\documentclass[nofootinbib,reprint,amsmath,amssymb,aps]{revtex4-1}
\usepackage[utf8,latin1]{inputenc}
\usepackage{graphicx}
\usepackage[dvipsnames]{xcolor}
\usepackage{dcolumn}
\usepackage{amssymb}

\usepackage{mathabx}
\usepackage{mathrsfs}  
\usepackage{tensor}
\usepackage[normalem]{ulem}
\usepackage{cancel}
\usepackage{bm}
\newcommand{\qm}[1]{``#1''}
\usepackage{hyperref}
\usepackage{stackengine,scalerel}
\hypersetup{colorlinks, linkcolor={red},citecolor={blue},urlcolor={blue}}

\newcommand{\dd}{{\rm d}}

\begin{document}

\title[Analytical coordinate time at the second post-Newtonian order]{Analytical coordinate time at the second post-Newtonian order}

\author{Vittorio De Falco$^{1,2}$}\email{v.defalco@ssmeridionale.it}
\author{Marco Gallo$^{3}$} \email{gallomarco1989@gmail.com}

\affiliation{
$^1$ Scuola Superiore Meridionale, Largo San Marcellino 10, 80138 Napoli, Italy,\\
$^2$ Istituto Nazionale di Fisica Nucleare, Sezione di Napoli, Complesso Universitario di Monte S. Angelo, Via Cintia Edificio 6, 80126 Napoli, Italy\\
$^3$ Ministero dell'Istruzione e del Merito (M.I.M., ex M.I.U.R.)
}

\date{\today}

\begin{abstract}
We derive the analytical expression of the coordinate time $t$ in terms of the eccentric anomaly $u$ at the second post-Newtonian order in General Relativity for a compact binary system moving on eccentric orbits. The parametrization of $t$ with $u$ permits to reduce at the minimum the presence of discontinuous trigonometric functions. This is helpful as they must be properly connected via accumulation functions to finally have a smooth coordinate time $t(u)$. Another difficulty relies on the presence of an infinite sum, about which we derive a compact form. This effort reveals to be extremely useful for application purposes. Indeed, we need to truncate the aforementioned sum to a certain finite threshold, which strongly depends on the selected parameter values and the accuracy error we would like to achieve. Thanks to our work, this analysis can be easily carried out.  
\end{abstract}
\maketitle

\section{Introduction}
General Relativity (GR) is the best theory of gravity so far available as it received numerous observational confirmations in the course of history and especially in the last decade \cite{Clifford2014,Ishak2019,Volkel2021,LIGO2021}. From an astrophysical perspective, the main targets of investigation are the compact binary systems, since they represent the natural laboratories for inquiring gravity. However, it is well known that GR possesses a non-linear geometric structure, which makes arduous not only the description of the compact binary system's dynamics, as it is ruled by retarded-partial-integro differential equations \cite{Maggiore:GWs_Vol1,Blanchet2014,Poisson-Will2014}, but also the subsequent benchmark against the observations.

A way out to deal with the tangled GR architecture is by exploiting the \emph{post-Newtonian (PN) method} \cite{Lorentz1937}. It assumes that the two bodies are slowly moving, weakly self-gravitating, and weakly stressed (also known as \emph{PN gravitational sources}) \cite{Maggiore:GWs_Vol1,Blanchet2014}, as well as largely separated. These hypotheses allow to treat the bodies as test particles and the curved background can be approximated as a Newtonian absolute Euclidean space on which we can add the relativistic corrections in power-series of $1/c$, where $n$PN order means to neglect terms higher than $1/c^{2n}$. This approach is advantageous, since it leads to ordinary differential equations of motion (EoMs), which still preserve their relativistic nature (i.e., invariance under a global PN-expanded Lorentz transformation) \cite{Blanchet2014,Poisson-Will2014}. Nevertheless, they spoil the GR general covariance in favour of particular coordinate systems \cite{Blanchet2014,Poisson-Will2014}.      

The PN approximation strategy finds disparate applications in the astrophysical context, like: relativistic two-body dynamics and gravitational back-reaction in binary pulsars \cite{Taylor:1989sw,Damour:1990wz,Kramer:2006nb,Kramer:2021jcw}; direct detection of gravitational waves (GWs) from coalescing compact binaries \cite{Blanchet2014}; precision tests of gravity theories \cite{Wex:2014abc}; neutron star mass measurements in binary pulsars \cite{Ozel:2016oaf};  inquiring the extreme mass ratio inspiral candidates and looking for their nature \cite{Seoane2018}. Depending on the sensibility of the observational data or also on the theoretical targets to achieve, different PN orders of binary system dynamics are employed.  

In this letter, we aim at deriving the analytical formula of the coordinate time $t$ at the 2PN order. In our previous work \cite{LetteraDBA}, we provided the 1PN-accurate analytical formula of $t(\varphi)$, where $\varphi$ is the polar angle. Here, we will determine $t(u)$ at the 2PN order, with $u$ being the eccentric anomaly. The two expressions can be linked via the transformation $\varphi=\varphi(u)$. At the 1PN order we used the analytical formula of the relative distance $R(\varphi)$ about the two-body dynamics derived by Damour and Deruelle \cite{Damour1985}, whereas here we employ the result due to Sch\"afer and Wex about $R(u)$ at the 2PN order \cite{SchaferWex-1993}.

This result finds a fruitful application in inspiralling black hole binaries, both for rapidly extracting information on the gravitational source under study \cite{Mingarelli2019}, but also for generating template matching in ground- and space-based GW astronomy to fit the observational data \cite{Schmidt2024}. Indeed, the gravitational signal requires on the $x$-axis the coordinate time and on the $y$-axis the strain. Therefore, having on both axes analytical functions, this permits to quickly produce this parametric plot in terms of $u$ or $\varphi$. 

This letter is organized as follows: in Sec. \ref{sec:preliminaries} we provide definitions and set out our notation; in Sec. \ref{sec:methodology} we describe the methodology to derive the analytical formula of the coordinate time at the 2PN order; finally in Sec. \ref{sec:final} we conclude by discussing the implication of our work.

\section{Preliminaries}
\label{sec:preliminaries}
We consider a compact binary system in eccentric orbit composed by two self-gravitating bodies with masses $m_1>m_2$, total mass $M=m_1+m_2$, position vectors $\bm{r_1}(t)$ and $\bm{r_2}(t)$ and, velocity vectors $\bm{v_1}(t)$ and $\bm{v_2}(t)$, where $t$ is the coordinate time. Moreover, we define the separation vector $\bm{R}(t)=\bm{r_1}(t)-\bm{r_2}(t)$ and its modulus $R=|\bm{R}(t)|$, the reduced mass $\mu=m_1 m_2/M$, the symmetric mass ratio $\nu=\mu/M$, and the unit vector $\bm{n}=\bm{R}(t)/R$. Adopting harmonic coordinates in the center-of-mass frame and defining $\bm{r}=\bm{R}/(GM)$ and $r=|\bm{r}|$, the EoMs at the 2PN order read as \cite{Memmesheimer-2004,SchaferWex-1993}: 
\begin{subequations} \label{eq:EoMs}
\begin{align}
{\dot r}^2&\equiv\frac{1}{s^4}\left( \frac{\dd s}{\dd t} \right)^2=\sum_{j=0}^{5} A_j s^j,\label{eq:EoM1}\\
\frac{\dd\varphi}{\dd s}&=\frac{\sum_{j=0}^{3} B_j s^j}{\sqrt{(s_{-}-s)(s-s_{+})}},\label{eq:EoM2}
\end{align}    
\end{subequations}
where $s=1/r$, $s_{-}$ and $s_{+}$ are the inverse of the apastron and periastron, respectively. The last quantities are obtained by searching for the non-vanishing positive roots and having finite limit as $\frac{1}{c}\rightarrow 0$ of the fifth-degree polynomial in Eq. \eqref{eq:EoM1}. The coefficients $A_j$ and $B_j$ are functions of $\nu$, total 2PN conserved energy $E=E_0+\frac{1}{c^2}E_1+\frac{1}{c^4}E_2+\mathcal{O}(c^{-6})$, and total reduced and conserved 2PN angular momentum $h=h_0+\frac{1}{c^2}h_1+\frac{1}{c^4}h_2+\mathcal{O}(c^{-6})$ with $h=J/(GM)$ and $J$ being the total conserved angular momentum. The explicit expressions of $E$ and $h$ up to the 2PN order can be found in Eqs. (23a) -- (23h) in Ref. \cite{Memmesheimer-2004}. The EoMs can be written by employing the following \qm{Keplerian-like} parametrization \cite{Memmesheimer-2004}:
\begin{subequations} \label{eq:QuasiKepPar}
\begin{align}
r&=a_r(1-e_r \cos{u}),\label{eq:QuasiKepPar1}\\
\frac{2\pi}{K}(\varphi-\varphi_0)&=v+\frac{f_{4\phi}}{c^4}\sin{(2v)}+\frac{g_{4\phi}}{c^4}\sin{(3v)},\label{eq:QuasiKepPar3}\\
v&=2\arctan{\left[\left(\frac{1+e_{\varphi}}{1-e_{\varphi}}\right)^{1/2}\tan{\frac{u}{2}}\right]},\label{eq:QuasiKepPar4}
\end{align}    
\end{subequations}
where $u$ and $v$ are the eccentric and true anomalies, respectively, $\varphi$ is the polar angle, $\varphi_0$ and $t_0$ are the initial orbital phase and initial time, respectively. In Eq. \eqref{eq:QuasiKepPar3}, the factor $\frac{2\pi}{K}$ gives the angle of advance of the periastron per orbital revolution, where $K$ is a function of $\nu, E, h$. Finally $a_r$ is the semi-major axis of the orbit, $e_r$, $e_{\varphi}$ the PN eccentricities, and $f_{4\varphi}$, $g_{4\varphi}$ some PN functions depending on $\nu, E, h$. The explicit expressions of all the aforementioned quantities can be found in Refs. \cite{Memmesheimer-2004,SchaferWex-1993}. Following our definitions, we commit a little abuse of notation, because the coordinate time $t$ scales as $t(GM)$
and another variable should be adopted \cite{Memmesheimer-2004}.

\section{Methodology}
\label{sec:methodology}
We aim at relating the coordinate time $t$ to the angle $u$ using the following relation
\begin{align}\label{eq:ODE1}
    \psi (s)&\equiv\frac{\dd\varphi}{\dd t}=\frac{\dd\varphi}{\dd s}\cdot\frac{\dd s}{\dd t}\notag\\
    &=s^2\frac{\left(\sum_{j=0}^{3} B_j s^j\right)\sqrt{\sum_{j=0}^{5} A_j s^j}}{\sqrt{(s_{-}-s)(s-s_{+})}}.
\end{align}
Starting from Eq. \eqref{eq:ODE1} and taking into account that the parameter $s=s(u)=1/r(u)$, we obtain
\begin{equation}\label{eq:ODE2}
    \dd t=\frac{1}{\psi (s(u))}\dd\varphi=\frac{1}{\psi (s(u))}\frac{\dd\varphi}{\dd u}\dd u\equiv \tau(u)\dd u,
\end{equation}
where $\tau(u)=\frac{1}{\psi (s(u))}\frac{\dd\varphi}{\dd u}$ and
\begin{align}\label{eq:ODE3}
    \frac{\dd\varphi}{\dd u}&=\left(\frac{2\pi}{K}\right)^{-1}\frac{\sqrt{1-{e_{\varphi}}^2}}{1-e_{\varphi}\cos u}\notag\\
    &\times\Biggr[1+2\frac{f_{4\varphi}}{c^4}\cos(2v)+3\frac{g_{4\varphi}}{c^4}\cos(3v)\Biggr].
\end{align}
Before to integrate Eq. \eqref{eq:ODE2}, we rewrite the quantities $\cos(2v)$ and $\cos(3v)$ in Eq. \eqref{eq:ODE3} in terms of an infinite power-series of $\cos(mu)$ as follows (see Appendix B in Ref. \cite{Boetzel-2017}, for their derivations and more details): 
\begin{subequations}\label{eq:vINu}
\begin{align}
      \cos(2v)&=\frac{2-{e_{\varphi}}^2-2\sqrt{1-{e_{\varphi}}^2}}{{e_{\varphi}}^2}+\frac{4\sqrt{1-{e_{\varphi}}^2}}{{e_{\varphi}}^2}\notag\\
      &\times\left[\sum_{m=1}^{+\infty}\beta^m \left(m\sqrt{1-{e_{\varphi}}^2}-1\right)\cos(mu)\right],\label{eq:vINu1}\\
      \cos(3v)&=\frac{3{e_{\varphi}}^2-4+(4-e_{\varphi}^2)\sqrt{1-{e_{\varphi}}^2}}{{e_{\phi}}^3}\notag\\
      &+\frac{2\sqrt{1-{e_{\varphi}}^2}}{{e_{\varphi}}^3}\left\{\sum_{m=1}^{+\infty}\beta^m 
      \left[2m^2(1-{e_{\varphi}}^2)\right.\right.\notag\\
      &\left.\left.-6m\sqrt{1-{e_{\varphi}}^2}+4-{e_{\varphi}}^2\right]\cos(mu)\right\},\label{eq:vINu2}    
\end{align}
\end{subequations}
where $\beta=(1-\sqrt{1-{e_{\varphi}}^2})/e_{\varphi}$. 

For applications, the infinite sum in Eq. \eqref{eq:vINu} must be truncated at a finite integer $m_{\rm max}\ge1$, chosen based on the desired approximation error we would like to achieve in the coordinate time formula. This threshold largely depends on the assigned parameter values. Consequently, the interested term in Eq. \eqref{eq:ODE3} can be expressed as
\begin{align}
\chi&=2\frac{f_{4\varphi}}{c^4}\cos(2v)+3\frac{g_{4\varphi}}{c^4}\cos(3v)\notag\\
&=\frac{1}{c^4}\Bigg{\{}
\sum _{m=1}^{m_{\rm max}} \Big[2 f_{4\varphi0} x(m)+3 g_{4\varphi} y(m)\Big]\cos(m u)\notag\\
&+2 f_{4\varphi0} x_0+3g_{4\varphi0}y_0\Bigg{\}}, \label{eq:TERM-INFINITE}
\end{align}
where $e_0=\sqrt{1+2 E_0 {J_0}^2}$ is the 0PN eccentricity, $e_1=h_0\sqrt{-E_0}$, and the other coefficients read as
\begin{subequations}\label{eq:compact-formula}
\begin{align}
f_{4\varphi0}&=\frac{e_0^2 [\nu  (19-3 \nu )+1]}{8 h_0^4},\\
g_{4\varphi0}&=\frac{e_0^3 (1-3 \nu ) \nu }{32 h_0^4},\\
x(m)&=\frac{4 e_1 \left(2 e_1 m-\sqrt{2}\right) \left(1-\sqrt{2} e_1\right)^m}{e_0^{m+2}},\\
y(m)&=\frac{ \left[2 e_1 \left(2 e_1 m^2-3 \sqrt{2} m+e_1\right)+3\right]}{2e_0\left(\sqrt{2} e_1 m-1\right)}x(m),\\
x_0&=\frac{2 e_1 \left(e_1-\sqrt{2}\right)+1}{e_0^2},\\
y_0&=\frac{e_0^3 \left[3 e_0^2+\sqrt{2} e_1 \left(2 e_1^2+3\right)-4\right]}{\left(1-2 e_1^2\right)^3}.
\end{align}    
\end{subequations}
Once we fix $\dot{r}(t_0)=0$ as initial condition, the function $\tau$ depends on the initial orbit eccentricity $\gamma$ (where $\dot{\varphi}(t_0)=\gamma\sqrt{GM/r_0^3}$ and $\gamma=\sqrt{1-e_0}$) and the initial separation $r_0=r(t_0)$ between the bodies. Performing some numerical simulations, we conclude that high eccentricities $e_0$ (or equivalently low initial orbit eccentricity $\gamma$) are responsible for the increment of the truncation error, whereas higher initial distances $r_0$ make it lower. 

The integrating function $\tau(u)$ can be split at the various PN orders as follows:
\begin{equation}\label{eq:IntFun}
     \tau(u)=\tau_{\rm 0PN}(u)+\tau_{\rm 1PN}(u)+\tau_{\rm 2PN}(u).
\end{equation}
The analytical formula of the coordinate time as a function of $u$ is obtained by solving the following integrals:
\begin{align}\label{eq:Int2}
   t(u)&=\int \Big{[}\tau_{\rm 0PN}(u)+\frac{1}{c^2}\tau_{\rm 1PN}(u)+\frac{1}{c^4}\tau_{\rm 2PN}(u)\Big{]} \,\dd u\notag\\
   &=t_{\rm 0PN}(u)+\frac{1}{c^2}t_{\rm 1PN}(u)+\frac{1}{c^4}t_{\rm 2PN}(u),
\end{align}
where we have set $t_0$ to zero without loss of generality. 

The numerical integration of Eq. \eqref{eq:Int2} gives a monotonically increasing function. The analytical solution $t(u)$ is obtained computing the three integrals in Eq. \eqref{eq:Int2}. The first two terms can be easily integrated, giving
\begin{subequations} \label{eq:TIME1PN}
\begin{align}
t_{\rm 0PN}(u)&=\frac{u-e_0 \sin u}{2 \sqrt{2} \left(-E_0\right)^{3/2}},\label{eq:SOL-0PN}\\
t_{\rm 1PN}(u)&=\frac{1}{8 \sqrt{2}e_0 (-E_0)^{5/2} }\biggl\{E_0^2 \biggl[8 h_0 h_1+4 e_1^2 (3 \nu -1)\notag\\
&-7 (\nu +1)\biggl]\sin u+2 e_1 \left(4 e_1^2-3\right)\sin u\notag\\
&+e_0 u \biggl[6 E_1-E_0^2 (\nu -15)\biggl]\biggl\}\label{eq:SOL-1PN}.
\end{align}
\end{subequations}
We checked that Eq. \eqref{eq:TIME1PN}, namely the 1PN order of the coordinate time, is equivalent to the 1PN formula provided in Ref. \cite{LetteraDBA}, written in terms of the polar angle $\varphi$ thanks to Eqs. \eqref{eq:QuasiKepPar3} and \eqref{eq:QuasiKepPar4} at the 1PN order. Our previous approach devised in Ref. \cite{LetteraDBA} gives rise to a discontinuous trigonometric function, whose periodic branches must be smoothly connected. In order to solve this issue, we introduce the concept of the \emph{accumulation function}, which permits to achieve the map regularity. Instead, the current parametrization of $t$ via the angle $u$ results to be more advantageous, as it avoids the aforementioned discontinuities, leading thus directly to a regular function. 

Finally, we need to determine only the 2PN contribution $t_{\rm 2PN}$, which can be written as
\begin{align}\label{eq:IntegrandFun2PN}
  \int \tau_{\rm 2PN}(u) \,\dd u&=\mathcal{C}\left[\sum_{i=0}^{3} a_{i}\mathcal{I}(i,1)+\sum_{i=0}^{4} b_{i}\mathcal{I}(i,2)\right.\notag\\
  &\left.+\sum_{i=0}^{5} c_{i}\mathcal{I}(i,3)+\sum_{i=2}^{m_{\rm max}} d_{i}\mathcal{J}(i)\right], 
\end{align}
where the coefficients $\mathcal{C},a_{i},b_{i},c_{i},d_{i}$ depend on $\nu, E, h$ and can be found in the \emph{Supplementary Materials}, whereas
\begin{subequations}\label{eq:Integral2PN_1}
\begin{align}
\mathcal{I}(n,m)&=\int \frac{\cos ^nu\,\dd u}{[-e_0+(1-2 e_1) \cos u] (1-e_0 \cos u)^m},\\
\mathcal{J}(n)&=\int \frac{\cos (nu)\,\dd u}{[-e_0+(1-2 e_1) \cos u] (1-e_0 \cos u)}.
\end{align}        
\end{subequations}
After the integration process, we obtain
\begin{equation}\label{eq:SOL-2PN}
    t_{\rm 2PN}(u)= D_1\mathcal{A}_1 +D_2\mathcal{A}_2 +D_3,
\end{equation}
where
\begin{subequations} \label{eq:DISC-FUNC}
\begin{align} 
\mathcal{A}_1&=\arctan{\left[\sqrt{\frac{1+e_0}{1-e_0}} \tan \left(\frac{u}{2}\right)\right]},\label{eq:coeff_SOL}\\
\mathcal{A}_2&=\arctan \left[\sqrt{\frac{e_0-2 e_1+1}{e_0+2 e_1-1}} \tan \left(\frac{u}{2}\right)\right];
\end{align}
\end{subequations}
whereas the coefficients $D_1,D_2,D_3$ are listed in the \emph{Supplementary Materials}, for the particular case of $m_{\rm max}=10$\footnote{The extension to a generic $m_{\rm max}$ is not presented, because it is difficult to achieve symbolically. However, thanks to Eq. \eqref{eq:compact-formula} it is possible to easily obtain the desired case.}. It is important to note that the functions $\mathcal{A}_1$ and $\mathcal{A}_2$ are both discontinuous, so we must smoothly connect the different branches through an accumulation function. It can be easily checked that $\mathcal{A}_1$ and $\mathcal{A}_2$ share the same accumulation function, which is
\begin{equation} \label{eq:CONT-FUNC}
F(u)=\begin{cases}
0 & \mbox{if}\ u\in[0,\pi],\\
\pi\left\{\left[\frac{u-\pi}{2 \pi }\right]+1\right\}& \mbox{otherwise},
\end{cases}    
\end{equation}
where the symbol $[\cdot]$ in the above expression represents the integer part. Therefore, the continuous functions are
\begin{align}\label{eq:CONT}
\bar{\mathcal{A}}_1&=\mathcal{A}_1+F(u),\qquad \bar{\mathcal{A}}_2=\mathcal{A}_2+F(u),
\end{align}
which must be substituted in Eq. \eqref{eq:SOL-2PN} to obtain the smooth coordinate time at the 2PN order.

\begin{figure}[ht!]
\includegraphics[scale = 0.3]{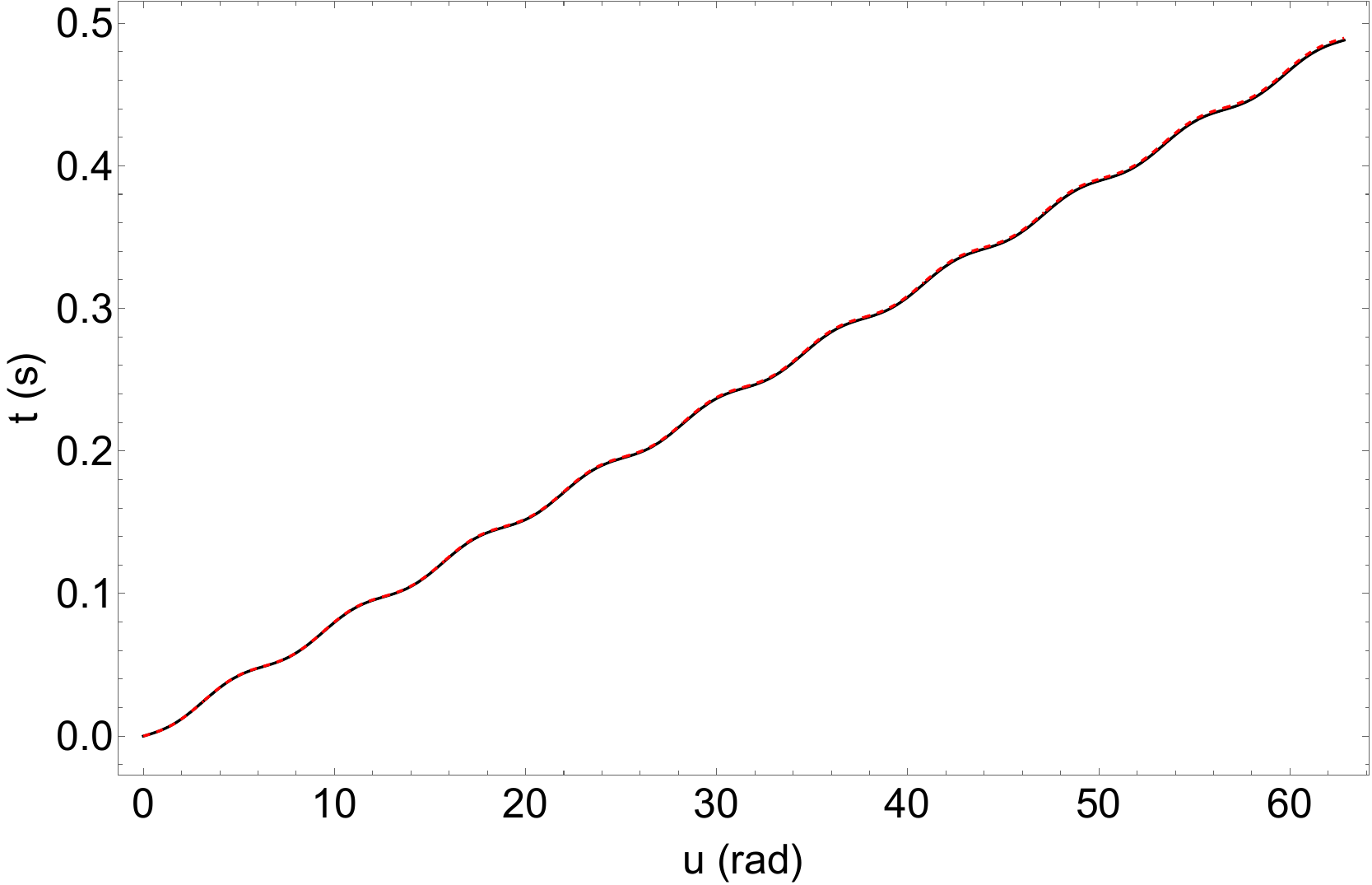}
\caption{Numerical integration of $t(u)$ (black line) and analytical formula \eqref{eq:Int2} (red dashed line) with $u\in[0,20\pi]$, where the coordinate time has been scaled by $GM$ (i.e., $t(GM)$). The following parameter values have been used: $m_1=1.60 M_{\odot}$, $m_2=1.17 M_{\odot}$, $\gamma=0.7$, $r_0(GM)=100M$, and $\dot{r}(0)=0$. The infinite sum has been truncated at $m_{\rm max}=10$. We have employed the same values exploited in Fig. 3 of Ref. \cite{LetteraDBA} for facilitating the comparison with the 1PN accurate formula.}
\centering
\label{fig:Fig1}
\end{figure}

In Fig. \ref{fig:Fig1} we see a good agreement between the numerical integration and the analytical formula of the coordinate time. The mean relative error\footnote{The relative error has been defined as the difference of the numerical and analytical expressions of the coordinate time in modulus over the absolute value of the numerical formula.} for our input parameters and selecting $m_{\rm max}=10$ amounts to $0.3\%$\footnote{We have chosen a high value of $m_{\rm max}$ just as an example for showing the power of our compact form \eqref{eq:TERM-INFINITE}.}.

\section{Discussion and conclusions}
\label{sec:final}
The analytical formula of the coordinate time (cf. Eqs. \eqref{eq:TIME1PN}, \eqref{eq:SOL-2PN}, \eqref{eq:CONT-FUNC}) at the 2PN order represents a new result in the GR PN literature. In our previous work, where we provided the analytical expression of the coordinate time at the 1PN order \cite{LetteraDBA}, we exploited and introduced the idea of accumulation function. Instead, here we have bypassed this concept at the 1PN order thanks to the parametrization of the coordinate time in terms of $u$. However, we have seen that the accumulation function becomes again crucial in the moment to work out the 2PN term, after the integration process of Eq. \eqref{eq:DISC-FUNC}. 

Another original facet with respect to the 1PN case is the presence of an infinite sum (cf. Eq. \eqref{eq:vINu}), which we arranged in a simpler and more compact form (cf. Eq. \eqref{eq:TERM-INFINITE}). This effort is convenient for all applications, as the series must be truncated at a certain finite $m_{\rm max}$, which can be determined once the parameter values are assigned and the approximation error is chosen. This preliminary analysis can be easily conducted via Eq. \eqref{eq:TERM-INFINITE}.

Our formula at the 2PN level can replace the numerical schemes implemented in the most used astrophysical template bank codes for generating gravitational waveforms during the inspiral stage \cite{Ajith2008}. Our proposal results to be of more flexible, simple, and fast implementation with reasonable and doable approximation costs. We underline that we have settled a strategy that revealed to be successful, because we have generalized it from the 1PN to the 2PN order. In addition, we have properly extended and refined it, as in the 2PN case we do not have a proper explicit solution of the radius in terms of an angle. In addition, there is not a simple relation linking $\varphi$ and $u$ (cf. Eq. \eqref{eq:QuasiKepPar3}), which leads then to Eq. \eqref{eq:vINu} involving the aforementioned infinite sum. In particular, the use of the parameter $u$ is preferred more than the polar angle $\varphi$, because it is naturally suggested from how the problem is formulated at the 2PN order (cf. Eq. \eqref{eq:QuasiKepPar}). 

This work establishes another fundamental tassel in the PN literature for mainly two reasons: (1) having a more accurate time coordinate formula; (2) reinforcing the methodology to deal with the coordinate time at higher PN orders. We underline that the illustrated procedure has never been advanced by other authors in the literature. Furthermore, our program to derive the analytical expression of $t$ at higher and higher PN orders permits also to: elaborate new refined approaches; understanding more about the PN effects on the coordinate time; figuring out new parametrization strategies, which might dodge the smooth junction business. 

This continuous need to attain such higher and higher PN levels theoretically is closely linked to the steady upgrade in the sensitivity of actual and near future observational instruments \cite{Blanchet2014,Blanchet2024}. This scientific evidence allows then to better investigate gravity and black hole physics. Therefore, our work configures more and more timely and prominent in this scenario. We underline that our formula is useful for computing the orbital period of compact binary systems during the inspiral stage and can be also exploited to fit the observational data to extract information from the gravitational sources under study.  

\acknowledgements
VDF is grateful to Gruppo Nazionale di Fisica Matematica of Istituto Nazionale di Alta Matematica for support. VDF acknowledges the support of INFN {\it sez. di Napoli}, {\it iniziative specifiche} TEONGRAV. The authors are grateful to Alessandro Ridolfi and Delphine Perrodin for the useful discussions and suggestions. 

\section*{Data availability}
No new data were generated or analysed in support of this research.

\bibliography{references}
   
\end{document}